\documentclass[sigconf]{acmart}

\usepackage{balance}

\def\BibTeX{{\rm B\kern-.05em{\sc i\kern-.025em b}\kern-.08emT\kern-.1667em\lower.7ex\hbox{E}\kern-.125emX}}

\usepackage{booktabs} 
\usepackage{tikz}
\usepackage{graphicx,epsfig,color,endnotes,alltt}
\usepackage{caption}
\usepackage{mathtools}
\DeclareCaptionType{copyrightbox}
\usepackage{comment}
\usepackage{keyval}
\usepackage[font=footnotesize,subrefformat=parens,labelformat=parens]{subfig}
\usepackage{url}
\usepackage[linesnumbered,ruled,vlined]{algorithm2e}
\usepackage{fancyvrb}
\usepackage[noindent, nolineno]{lgrind}
\usepackage{epstopdf}
\usepackage{array}
\usepackage{multirow,makecell}
\usepackage{amsmath}

\usepackage{bm}
\usepackage{slashbox}
\usepackage[labelfont=bf, justification=justified]{caption}
\makeatletter 
\newcommand\notsotiny{\@setfontsize\notsotiny\@vipt\@viipt}
\makeatother

\usepackage{xcolor}
\usepackage{listings}
\definecolor{dkgreen}{rgb}{0,0.6,0}
\definecolor{gray}{rgb}{0.5,0.5,0.5}
\definecolor{mauve}{rgb}{0.58,0,0.82}

\SetKwInOut{Input}{Input}
\SetKwInOut{Output}{Output}
\SetKwInOut{Require}{Require}

\usepackage[titletoc,title]{appendix}

\definecolor{dgreen}{rgb}{0.00, 0.75, 0.00}

\begin{document}
\fancyhead{}
\title{Optimizing Memory Performance of Xilinx FPGAs under Vitis}
\author{Ruoshi Li}
\affiliation{%
	\institution{Services Computing Technology and System Lab \\
		Cluster and Grid Computing Lab \\
		School of Computer Science and Technology \\
		Huazhong University of Science and Technology}
	\city{Wuhan}
	\state{430074}
	\country{China}}
\email{liruoshi3@hust.edu.cn}

\author{Hongjing Huang,Zeke Wang}
\affiliation{%
	\institution{
		Zhejiang University}
	\city{Zhejiang}
	\state{310058}
	\country{China}}
\email{21515069@zju.edu.cn, wangzeke@zju.edu.cn}

\author{Zhiyuan Shao,Xiaofei Liao,Hai Jin}
\affiliation{%
	\institution{Services Computing Technology and System Lab \\
		Cluster and Grid Computing Lab \\
		School of Computer Science and Technology \\
		Huazhong University of Science and Technology}
	\city{Wuhan}
	\state{430074}
	\country{China}}
\email{zyshao@hust.edu.cn,xfliao@hust.edu.cn, hjin@hust.edu.cn}



\begin{abstract}
Plenty of research efforts have been devoted to FPGA-based acceleration, due to its low latency and high energy efficiency. However, using the original low-level hardware description languages like Verilog to program FPGAs requires generally good knowledge of hardware design details and hand-on experiences. Fortunately, the FPGA community intends to address this low programmability issues. For example, the FPGA vendor Xilinx has provided Vitis to reduce programming efforts needed by FPGA programmers who intends to leverage FPGA’s computing power, with the intention that programming FPGAs is just as easy as programming GPUs. Even though Vitis is proven to increase programmability, we cannot directly obtain high performance without careful design regarding hardware pipeline and memory subsystem. In this paper, we focus on the memory subsystem, comprehensively and systematically benchmarking the effect of optimization methods on memory performance. Upon benchmarking, we quantitatively analyze the typical memory access patterns for a broad range of applications, including AI, HPC, and database. Further, we also provide the corresponding optimization direction for each memory access pattern so as to improve overall performance.   

\end{abstract}





\maketitle

\vspace{-1ex}
\section{Introduction}

With the development of computer architecture, there is a considerable gap between the higher performance of the computing units and the slower speed of DRAM memory systems. With the development of various applications, such as Neural Network training that require large-scale data exchange, several research institutes such as Samsung and Micron have presented next-generation high-performance memory architectures, like Hybrid Memory Cube (HMC)~\cite{jeddeloh2012hybrid} and High Bandwidth Memory (HBM)~\cite{jun2017hbm}. In this paper, we intend to optimize memory performance on FPGAs~\cite{XilinxU280:2020} with High-Level Synthesis(HLS)~\cite{XilinxHLS:2020}. 

$\bullet$ The provided max bandwidth implementation by HLS in HBM is up to 431GB/s~\cite{XilinxHBM:2020}, which achieves the identical performance with the implementation by Register-Transfer Level (RTL)~\cite{XilinxRTL:2009}, such as Verilog or System Verilog.

$\bullet$For the latency of HBM, the HLS implementation based on the Vitis platform~\cite{XilinxVitis:2020} and the RTL implementation have the same latency for HBM data access. Compared with the memory access delay of DDR4~\cite{o2017fine,Micron:2015,mi2010software}, the HBM access latency is more extensive. The increasing latency is because the connection between the on-chip IO and HBM memory structure is a crossbar controller~\cite{XilinxHBMFPGA:2019}. Compared with the DDR controller, its structure causes the average latency to increase. Compared with the implementation of the Vivado platform, the latency based on the Vitis platform has an absolute increase. The reason is that Vitis will encapsulate the kernel, whether implemented by HLS or RTL, which leads to an increase in latency.

$\bullet$The Vitis-based HLS implementation of HBM access is equivalent to the implementation of CPU/GPU access to HBM. because of implemented on FPGA structure, it close to the HBM connection without multiple cache and control structure interference, and HLS implementation The HBM memory access parameters are the same as the direct access to the AXI port of HBM. Therefore, using the characteristics of the HLS and Vitis platforms can be similar to the CPU/GPU architecture~\cite{XilinxUltraFast:2020}. By benchmarking many fundamental parameters, we obtain the performance of  HBM  under different implementations, which provides a benchmark for various future applications with various access modes on the HBM platform.

$\bullet$The Address Mapping Policy is Critical to High Bandwidth. Different address mapping policies lead to an order of magnitude throughput differences when running a typical memory access pattern (i.e., sequential traversal) on HBM, indicating the importance of matching the address mapping policy to a particular application.

With the development of FPGA applications, numerous high-concurrency and high-performance applications implemented on FPGAs, such as AI, HPC, graph computing, have enormous demands such as ease to use and data storage access performance on FPGA. With the emergence of a generation of high-performance storage structures, such as HBM and HMC, how to efficiently use FPGAs to handle these high-performance storage has become an important topic. With Xilinx launching multiple platforms U280, U250, and Vitis based on Vitis U200, U50, and Xilinx Run Time (XRT)~\cite{xrt:2019} based on HLS and OpenCL further make it possible to implement high-performance hardware-based on High-level programming language. In order to make better use of High-level programming language to bring out the performance of the next-generation memory, the paper will be based on the U280 platform, using the HLS implementation based on the XRT and Vitis platforms, and fully implement the FPGA's HBM memory access performance and features in the XRT environment. A wide range of benchmark compared with the primary memory access benchmark based on the RTL implementation on the Vitis platform and the RTL implementation based on the standard Vivado platform and the DDR memory access benchmark. Through these large-scale and comprehensive benchmark, we will have a complete understanding of the HBM memory access performance and characteristics of FPGAs based on High-level programming language, and through the ease of use achieved by HLS, the benchmark architecture can also provide a universal benchmark platform.


\section{Background}

We now present the preliminaries for understanding Vitis \cite{XilinxVitis:2020} and memory on FPGAs.

FPGA, with the arrival of the next golden decade of architecture, its application becomes more and more extensive such as AI and HPC\cite{zhang2017improving,turkington2006fpga,wei2017automated,zhang2017frequency,chen2016accelerating}, as well as database\cite{arcas2014empirical,owaida2017centaur,mueller2009fpga} and graph-related\cite{shao2019improving,Khoram:2018,Zhang:2018} algorithms; most of these algorithms have complex implementations and need to read and write a large amount of data. Hence, high-level languages such as High-Level Synthesis(HLS) and OpenCL in FPGA are the primary implementation method to simplify and migrate AI and other algorithms\cite{zhang2017improving,turkington2006fpga,gautier2016spector,cong2011high,zhou2018rosetta,hara2008chstone,muslim2017efficient,kim2017heterogeneous,de2018designing, neuendorffer2013building,daoud2014survey,alias2013optimizing}. At the same time, with the realization of such algorithms, standard DDR-based FPGAs have gradually entered performance bottlenecks\cite{sohi1991high,mi2010software,o2017fine,panda1998incorporating}. 

With the next generation of storage showing outstanding performance and characteristics on the CPU and GPU\cite{manegold2002generic}, the HMC implementation in FPGA has demonstrated excellent performance in graph-related algorithms\cite{Zhang:2018,Khoram:2018}, and HBM is now the strength of FPGA High-performance off-chip memory\cite{du2020high}. Samsung has presented HBM, which is on the basis of DRAM 2.5D stacking and package technology and multiple DRAM dies interconnected through Through Silicon Via (TSV)\cite{jun2017hbm}. Take the FPGA-based HBM architecture implemented by Xilinx as an example\cite{alyushin2018bit}. It utilizes 2 HBM Stacks to embed its memory control system into the logic block of the FPGA. The memory controller contains 16 memory channels and 16 memory channels. The 32 pseudo channels expand to 32 AXI channels from the FPGA logic implementation\cite{XilinxAXI:2017,ArmAXI:2017}.

In terms of tools, to make FPGA implementation closer to CUDA, MPI, and other GPU and CPU implementations, Xilinx designed the Vitis platform and the Xilinx Runtime (XRT)\cite{xrt:2019} that supports the platform to integrate RTL, HLS, and OpenCL\cite{XilinxVitis:2020,XilinxUltraFast:2020}. The Vitis platform has adapted optimistic accessing memory modules such as HBM to the RTL design and high-level language implementations\cite{XilinxAlveo:2020,XilinxVitis:2020,XilinxHBMFPGA:2019}. This method has achieved perfect results in AI and HPC. In addition, XRT provides an interactive subsystem implemented by HOST and kernel. This system simplifies the implementation of PCIe and DMA, making it more convenient for the host to call FPGA, and its interactive efficiency is very close to that of GPU calls implemented by CUDA\cite{cong2018understanding}. There have even been several times of efficiency improvements in some AI-related implementations.


\section{design}
The goal of this section is to use Vitis to benchmark memory in terms of throughput and latency. In particular, we intend to measure the performance of the HLS implementation with various optimization parameters in Vitis. Therefore, our design and implementation mainly have two aspects. The first one is the memory access pattern based on the standard for loop, which is mainly to test some standard memory accesses' highest performance and the performance impact of some standard parameters. The second one is the memory access mode based on Dataflow, closer to the RTL method. It uses inter-function parallelism to decompose the memory access behavior of the standard for loop into address generation, memory access, and data operation, such that their memory accesses are completely independent. To this end, we can use the memory access behavior independence and function parallelism to measure memory access latency, some particular address behaviors (such as random addresses), and memory access based on HLS in discontinuous memory access features and performance.


In the following, we will first present a memory latency benchmarking engine that allows us to accurately measure memory latency with Vitis, followed by a memory throughout benchmarking engine that is used to measure memory throughput under different optimizations. 
Table~\ref{tab:param_label} illustrates all the symbols and their corresponding meanings used in this paper.
\begin{table}[htbp]
    \centering
    \caption{SUMMARY OF RUNTIME PARAMETERS}
    \label{tab:param_label}
    \begin{tabular}{|c||p{180pt}|}
        \hline
        Parameter&Definition \\
        \hline
        \hline
        $F$&Kernel frequency\\
        \hline
        $BW$&Memory bandwidth \\
        \hline
        $N$&Number of memory channels (normally $N=32$)\\
        \hline
        $W$&Bit-width of a memory transaction\\
        \hline
        $B$&Burst size\\
        \hline
        $NO$&Number of outstanding memory transactions\\
        \hline
        $T$& Run-time\\
        \hline
        $T_l$&Latency of one memory transaction\\
        \hline
        ${\tau}_{II}$&Iteration interval in a loop,or relative latency of memory transactions\\
        \hline
        $n$&No. of memory channels, $0\leq  n\geq31$\\
        \hline
        $i$&An iteration of a loop\\
        \hline
        $T_{s_i}$&Starting time of an memory transaction in an iteration of a loop \\
        \hline
        $T_{e_i}$&Ending time of an memory transaction in an iteration of a loop \\
        \hline
        $T_o$& Latency of an operation except memory transaction in a loop \\
        \hline
        $I$ & Total number of transactions\\
        \hline
        $FIFO$ & First in first out tunnel\\
        \hline
        $H_n$ & NO.n of memory tunnels\\ 
        \hline
        $G$ & Work group size of a memory channel\\
        \hline
    \end{tabular}
\end{table}
\subsection{Memory Latency Benchmark}
When design the memory latency benchmarking engine, we mainly address the following two challenges.

First, the memory latency benchmarking engine is implemented with Vitis and Xilinx Runtime (XRT), which abstract away the implementing details of memory transaction such that we are not able to configure each memory transaction whose interface is AXI in a fine-grained manner like implemented with Verilog~\cite{}.

\begin{figure}[htbp]
	\captionsetup[subfigure]{justification=centering}
    \centering
	\setlength{\abovecaptionskip}{+5pt}
	\setlength{\belowcaptionskip}{-15pt}

    \subfloat[Latency Benchmark Kernel  Architecture \phantom{1} ]{%
      \includegraphics[width=3in]{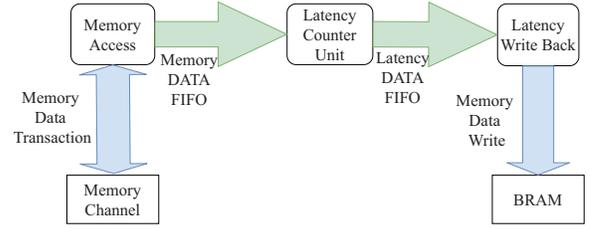}%
      \label{fig:latencys_archi}%
    }
    
    \subfloat[ Memory Access Unit \phantom{ 1} ]{%
      \includegraphics[width=3in]{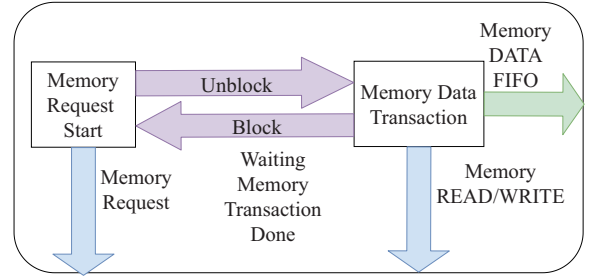}%
      \label{fig:latencys1_access}%
    }
    
    \subfloat[ Latency Counter Unit \phantom{1} ]{%
      \includegraphics[width=2in]{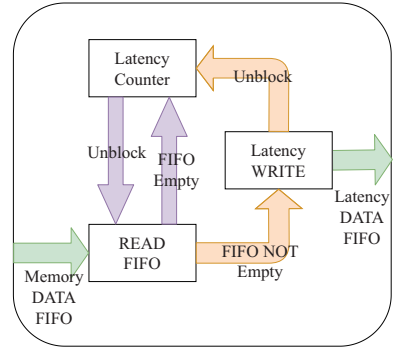}%
      \label{fig:latencys_counter}%
    }
    \caption{Hardware design of memory latency benchmark engine}
    \label{fig:latency_t_archi}
\end{figure}
Second, the framework built on Vitis and HLS is not suitable for direct use for simplicity and ease of use, compared to the Verilog-based memory benchmarking tool Shuhai~\cite{} that allows to directly send the latency numbers back to the host using the PCIe module. 

To address the above two challenges, we propose a memory benchmark architecture with HLS implementation. As shown in Figure~\ref{fig:latency_t_archi}, because HLS shields the timing details, we need to build a module that builds an independent accumulation loop structure. This module uses a loop accumulator and controls the loop in every clock cycle, accumulating and using it for sequence counting. For the memory access part of memory, as shown in Figure~\ref{fig:latencys1_access} and Algorithm~\ref{alg:memory access}, we have designed a module to perform a cyclic memory access. This cycle uses the principle of blocking loops to ensure that the module is blocked at the beginning of each memory access operation until the data returns so that we can obtain the clock delay of the memory accessing through each blocked memory access operation. However, to count each memory access operation's latency, as shown in Algorithm~\ref{alg:Latency count} and Figure~\ref{fig:latencys_counter}, we also need to parallel the memory access module with the loop accumulation module. We use the HLS Data Flow structure to construct these two modules and use a FIFO called memory data FIFO to link and control. Whenever the memory access module receives data, it writes the data into the memory data FIFO, and whenever the loop accumulator reads data from the FIFO, it clears the loop accumulator. Zero and transfer the accumulated value (as the latency period between two memory accesses) through another FIFO (latency data FIFO) to the last module. The last module writes data back, as shown in Algorithm~\ref{alg:write_back}. It will write each read from the latency data FIFO to another memory channel. Compared to Shuhai, the latency benchmark implemented on the Vitis module can only exchange data with the host through memory mapping. So in order to make HLS easy to use and complete data, we build a write-back module that writes the latency data read from the latency data FIFO back to another memory channel, ensuring data integrity while avoiding the latency data being disturbed by writing back to memory.

However, the HLS implementation involves two dependency memory channels (such as off-chip memory and another on-chip memory). The latency data obtained from one memory channel will be written back to another memory channel, due to the channel's memory access latency, which causes the cyclic accumulator to be blocked and generates incorrect latency statistics data. Therefore, to prevent this factor's influence, we need to ensure that the FIFO's length and the writing outstanding of the memory channel for writing latency data are far greater than the clock cycle of the memory access latency.

In addition to the HLS-based implementation, we also implement an RTL-based memory latency benchmark engine. This implementation is used on the Vitis platform as a benchmark to compare with the HLS implementation. The RTL implementation is based on the hardware implementation of Shuiai's C1 part\cite{wang2020shuhai}, and in order to adapt to the Vitis platform, we also write back the Shuiai's return data to another memory channel to prevent the whole hardware design from being optimized away.

\begin{algorithm}[htbp]
    \caption{Memory Accessing}
    \Input{The total transaction $I$, memory data channel ${H_n}_1$}
    \Output{memory data FIFO tunnel ${FIFO}_d$}
    \label{alg:memory access}
    \ForEach{$i \in [0,I-1]$}{
        $Result\gets {H_n}_1[i]$;\\
        ${{FIFO}_d}.Write\left(Result\right)$;
    }
\end{algorithm}
\begin{algorithm}[htbp]
    \caption{Latency count}
    \Input{The total transaction $I$,memory data FIFO tunnel ${FIFO}_d$ ,the latency counter $C$ }
    \Output{Latency data FIFO tunnel ${FIFO}_l$}
    \label{alg:Latency count}
    $C\gets 0$;\\
    \While{$i<I$}{
        \If{${FIFO}_d$ is not empty}{
            $Result \gets {FIFO}_d.read()$
            ${{FIFO}_l}.Write\left(C\right)$;\\
            $C\gets 0$;\\
            $i\gets i+1$;
        }
        \Else{
            $C\gets C+1$;
        }
    }
\end{algorithm}
\begin{algorithm}[htbp]
    \caption{Latency data write back to memory}
    \Input{The total transaction $I$, \\ Latency data FIFO tunnel ${FIFO}_l$}
    \Output{memory data channel ${H_n}_2$}
    \label{alg:write_back}
    \ForEach{$i \in [0,I-1]$}{
        $Latency\gets {FIFO}_l.read()$;\\
        ${H_n}_2[i]\gets Latency$;
    }
\end{algorithm}
The latency is on the basis of the memory hardware structure. In the above, we define the latency benchmark as the clock cycle from the beginning of a memory operation to the end of the memory operation.

\begin{equation}
\begin{aligned}
T_l = T_{s_{i+1}}-T_{e_i}
\end{aligned}
\label{equation:absolute latency}
\end{equation}
This memory access latency is absolute, as shown in Equation~\ref{equation:absolute latency}, and it comes from the physical implementation of memory and its controller. 

\subsection{Memory Bandwidth Benchmarking Engine}
In order to realize the benchmark influenced by multiple parameters of HLS, we need to consider the behavior of the high-level language under different parameters and its actual implementation on FPGA after compilation. For the implementation of HLS, we need to consider one main memory access feature: whether to fetch memory continuously and boundedly.

Continuously bounded memory access refers to whether there is a specific starting address during the memory access process. At the same time, after each round of memory access operation, its operations are independent, that is, the next round of memory access operations has nothing to do with the previous round.

For the continuous and bounded memory access implementation, we use the standard for loop to implement it.

\begin{figure}[t]
	\centering
	\setlength{\abovecaptionskip}{+5pt}
	\setlength{\belowcaptionskip}{-15pt}
	\includegraphics[width = 3in]{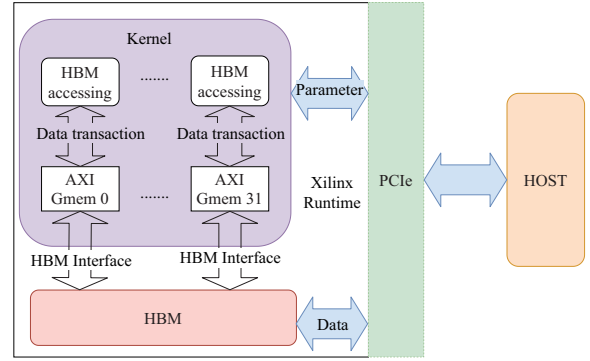}	
	\caption{Bandwidth Benchmark Architecture}
	\label{fig:benchmark}
\end{figure}

We adopt the dataflow method for the realization of non-continuous and bounded memory access, which is similar to the latency test method. We separate the memory access address generation and the memory access behavior and use the FIFO to control the memory access behavior and memory access boundary, thus forming an independent memory access module, such as the kernel in Figure~\ref{fig:benchmark}. Using 32 kernels for memory access and host timing, we can obtain the actual bandwidth and performance of 32 channels under different parameters.

There are two types of continuous memory accesses. One is the address continuous, and the other is the memory access request. Address continuity refers to an utterly continuous address generated during the memory access process without interruption. At this time, we can use the burst feature of AXI. In the case of an AXI handshake, we can directly read the burst size of memory. The continuous memory access request refers to continuously sending memory access requests to the memory controller, but the address is not necessarily continuous. At this time, there is a complete AXI handshake in each time memory accesses.

Bounded memory access refers to the existence of an exact starting address during the memory access process, such as a standard for loop.

Generally speaking, for HLS-based memory access implementations, continuous bounded memory access can achieve the highest performance. When an HLS implementation cannot achieve continuous bounded memory access, the actual impact on memory access performance is the outstanding parameter. All memory data accesses used at this time will carry out a complete AXI handshake protocol.

For actual HLS, a memory access operation latency II in the loop is relative. We call it a relative latency. This latency can be under the influence of actual implementation. The definition of this latency is from The clock cycle from the starting of memory fetch to the starting of the next memory fetch.

\begin{figure}[t]
\captionsetup[subfigure]{justification=centering}
\centering

\subfloat[ \phantom{H} ]{%
  \includegraphics[width=3in]{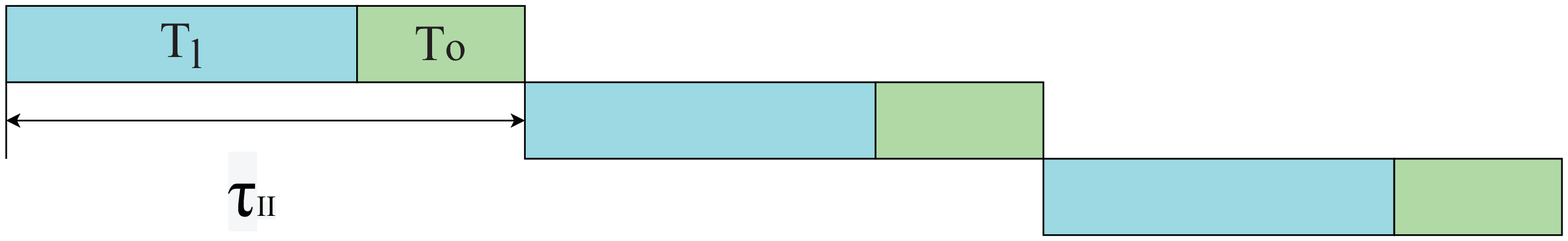}%
  \label{fig:rellatency_1}%
}

\subfloat[ \phantom{H} ]{%
  \includegraphics[width=3in]{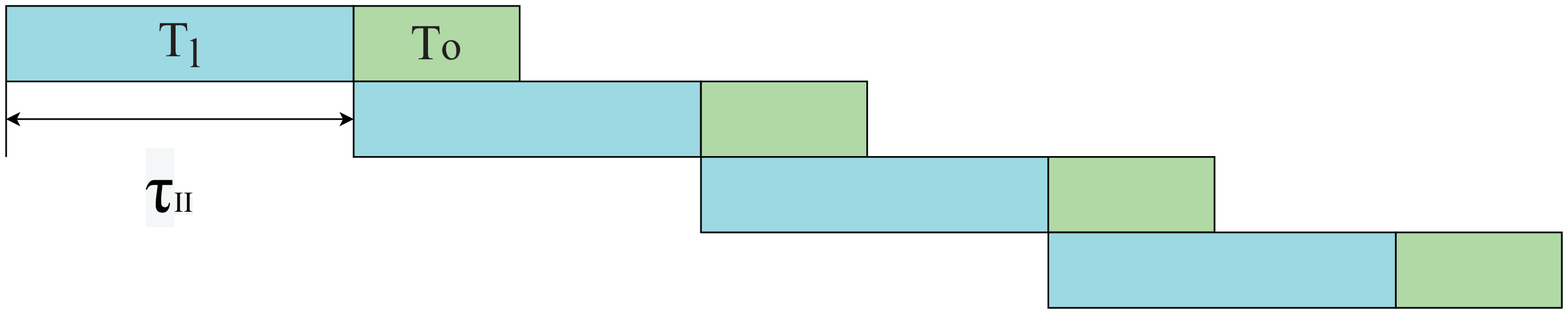}%
  \label{fig:rellatency_2}%
}

\subfloat[ \phantom{H} ]{%
  \includegraphics[width=3in]{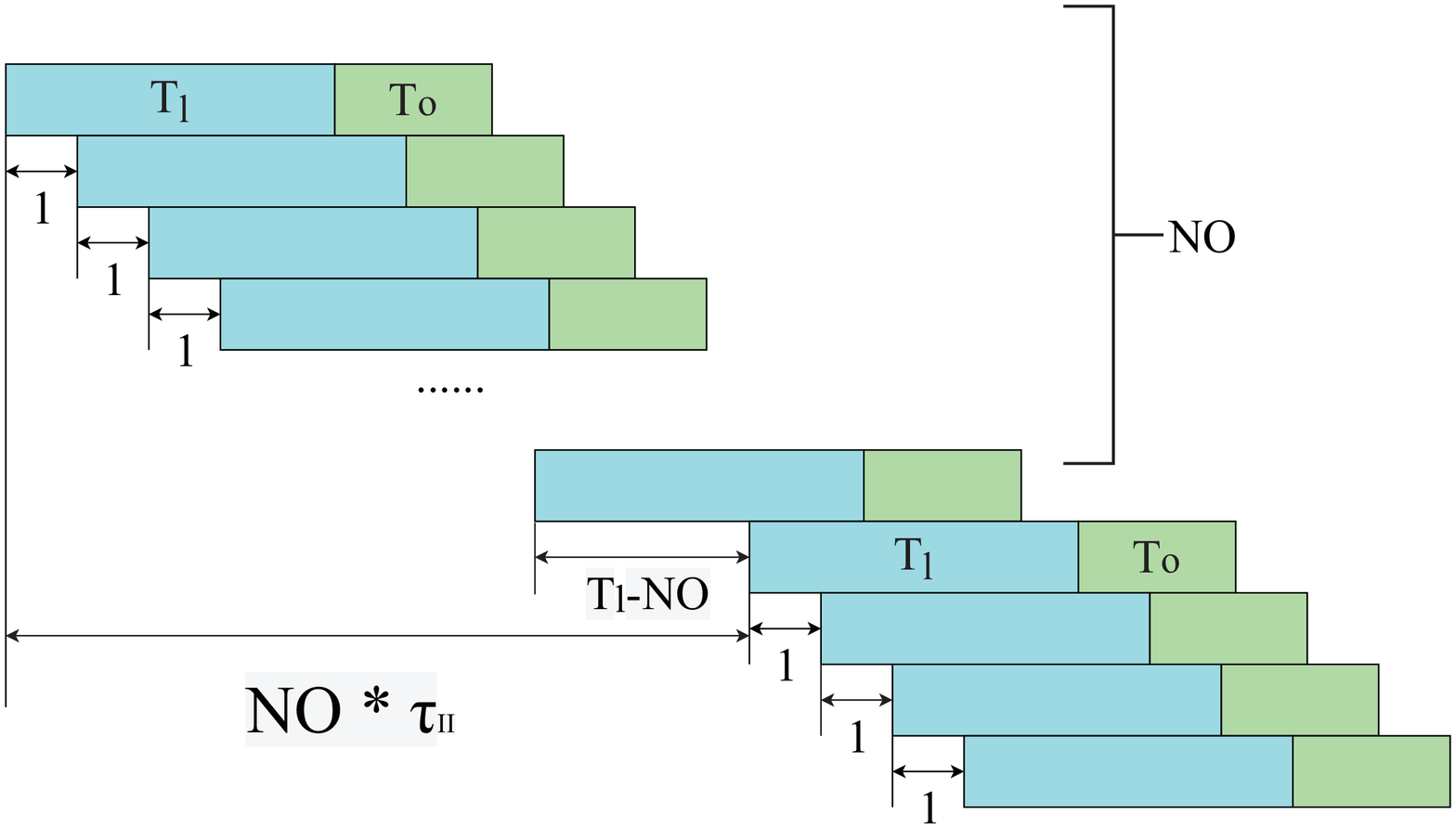}%
  \label{fig:rellatency_3}%
}

\caption{Relative Latency}
\label{fig:rellatency}
\end{figure}
\begin{equation}
\begin{aligned}
{\tau}_{II}=T_{s_{i+1}}-T_{s_i}=T_l+T_o
\end{aligned}
\label{equation:relative latency 1}
\end{equation}
The memory access latency and the absolute memory access latency, the relative access latency is composed of the total memory access latency, the number of memory access channels outstanding, and the operation latency after the memory access and the correlation of the data access (RAR, RAW, WAR, WAW).

As shown in Figure~\ref{fig:rellatency_1} and Equation~\ref{equation:relative latency 1}, in this case, the next memory access behavior must be performed after all operations in the previous cycle are completed, and the relative access latency is the highest at this time.

\begin{equation}
\begin{aligned}
{\tau}_{II}=T_{s_{i+1}}-T_{e_i}=T_l
\end{aligned}
\label{equation:relative latency 2}
\end{equation}
This access latency occurs in a completely no optimized FOR loop, or there is a correlation between the operation and the fetched data after the fetch (there is a correlation between cycles).

As shown in Figure~\ref{fig:rellatency_2}, when there is no correlation between the memory access operations, the PIPELINE optimization feature is used, and the next memory access behavior can be executed before the last operation completed, that is, the next memory access behavior is completed in the previous memory access behavior Then, the relative latency is illustrated in Equation~\ref{equation:relative latency 2}.

\begin{equation}
\begin{aligned}
{\tau}_{II}=MAX\left(1,\frac{N_O+T_l-N_O}{N_O}\right)
\end{aligned}
\label{equation:relative latency 3}
\end{equation}
fetched memory access data. The operation is only related to the current round of fetched data, but the fetched data is RAW or WAR.

As shown in Figure~\ref{fig:rellatency_3}, when there is no correlation between the operation and the fetched data after the fetch, the pipeline optimization feature is used, and the next fetch behavior can be executed when the previous fetch behavior incomplete. It is related to the critical parameter of an HLS implementation of memory access is outstanding. When there is an outstanding memory access cache, the AXI protocol used when accessing the memory can cache part of the data before memory accessing returns. When the number of outstanding cache channels is $NO$, the relative latency at this time is Equation~\ref{equation:relative latency 3}.

\section{Benchmarking}

When benchmarking memory, we plan to experiment with the throughput and latency under possible application scenarios. We need to carefully design the benchmark code to precisely show the effect of each parameter. 

\begin{figure}[htbp]
\captionsetup[subfigure]{justification=centering}
\centering

\subfloat[ HLS\phantom{ }]{%
  \includegraphics[width=2.3in]{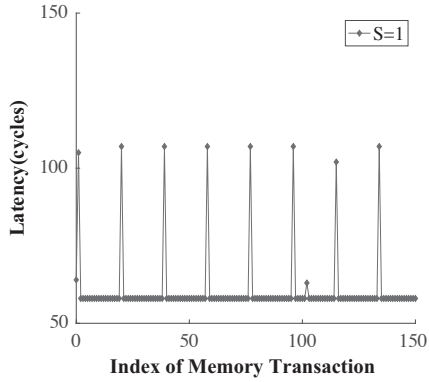}%
  \label{fig:latencys1_HLS}%
}

\subfloat[ RTL\phantom{ }]{%
  \includegraphics[width=2.3in]{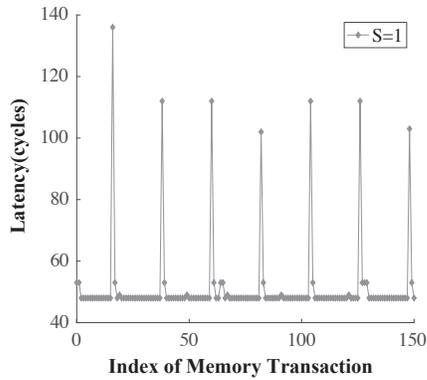}%
  \label{fig:latencys1_RTL}%
}

\subfloat[ DDR\phantom{ }]{%
  \includegraphics[width=2.3in]{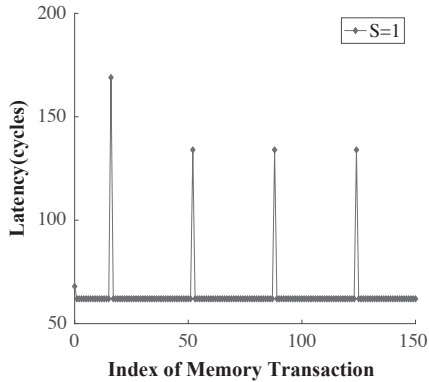}%
  \label{fig:latencys1_ddr}%
}

\caption{Latency of Various Implementation}
\label{fig:latencys_Imp}
\end{figure}

We can compare the performance between Vitis and Verilog to show the performance cost when using Vitis, which brings high programmability. 

The actual performance of memory access built on HBM and DDR, since each channel of the HBM memory access benchmark architecture implements on HLS, can point to an independent kernel. Assuming that the bit width of the $i$ channel is $W_i$, the memory access round Is $I_i$, the clock frequency is $F$, and the system running time (stated by Host) is $T$, then the actual memory access bandwidth achieved by HLS can be obtained.the Equation~\ref{equation:system bandwidth} shows system bandwidth.

\begin{equation}
\begin{aligned}
BW = 
\frac{\sum_{i=0}^{N-1}I_n * W_n}{T*8*10^9}
\end{aligned}
\label{equation:system bandwidth}
\end{equation}

In order to compare the deviation between HLS implementation and theoretical performance, we also need to derive the theoretical memory access bandwidth. Since AXI protocol is the actual implementation of memory access on FPGA which and each pair of AXI buses has at most one transmission request per clock, according to the HBM maximum, When bit width $W=256$ and all $N=32$ channels are activated automatically, the Equation~\ref{equation:theoretical bandwidth} shows the origin of theoretical bandwidth.

\begin{equation}
\begin{aligned}
BW = \frac{N*W*F}{8*10^9}
\end{aligned}
\label{equation:theoretical bandwidth}
\end{equation}

\subsection{Latency Testing}

The memory access latency accurately of consecutive memory read transactions when the memory controller is in an “idle” state, i.e., where no other pending memory transactions exist in the memory controller such that the memory controller can return the requested data to the read transaction with minimum latency. We aim to identify latency cycles of page hit, page closed, and page miss.

The “page closed” state occurs when a memory transaction accesses a row whose corresponding bank is closed, so the row Activate command is required before the column access.

The “page miss” state occurs when a memory transaction accesses a row that does not match the active row in a bank, so one Precharge command and one Activate command are issued before the column access, resulting in maximum latency.

The “page hit” state occurs when a memory transaction accesses a row that is open in its bank, so no Precharge and Activate commands are required before the column access, resulting in minimum latency.

\begin{figure}[h]
	\centering
    \includegraphics[width = 2.5in]{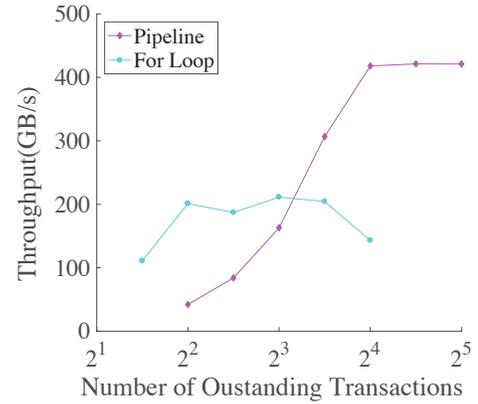}

	\caption{Effect of the number of outstanding channels}
	\label{fig:outstanding}
\end{figure}

As shown in Figure~\ref{fig:latencys_Imp}, to compare the impact of different implementations on memory access, this paper implements three different models of memory access latency benchmark.

First, as shown in Figure~\ref{fig:latencys1_HLS}, this implementation is based on the HLS under the Vitis architecture, which mainly performs latency benchmark on the HBM storage architecture.

The second, as shown in Figure~\ref{fig:latencys1_RTL}, is under the Vitis-based RTL implementation, and the latency benchmark is also performed on HBM. The last one, as shown in Figure~\ref{fig:latencys1_ddr}. The implementation is the same as the first one, but it is a latency benchmark for DDR memory architecture.

Through these three tests, the memory access latency under the HLS implementation is relatively close. Because the memory access structure in the HLS has the Gmem and the controller's cache layer, and there is a pre-judgment of the memory access behavior in the HLS. As a register buffer group, the delay based on HLS will be more significant than RTL, while the delay based on RTL will not. At the same time, DDR memory access latency is more stable than HBM, and the number of times high latency is less than HBM, which means that its page and bank are much more extensive than HBM.

\begin{figure}[htbp]
	\centering
	\setlength{\abovecaptionskip}{+5pt}
	\setlength{\belowcaptionskip}{-15pt}
    \includegraphics[width = 2.5in]{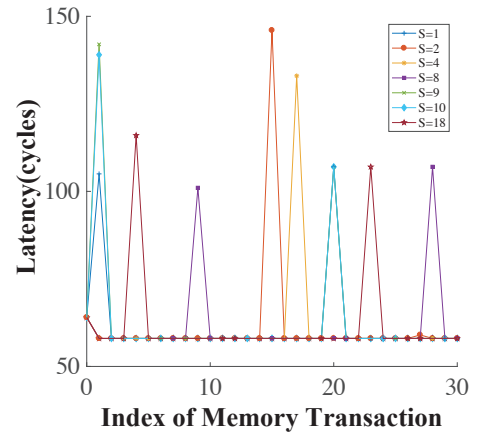}

	\caption{Memory Access Latency With Various Stride}
	\label{fig:latencys_str}
\end{figure}

As shown in Figure~\ref{fig:latencys_str}, this article uses different strides to test the page behavior of latency. Compared with the RTL implementation, the unit size of the data structure of the memory access pointer during implementation determines the step size of the HLS-based implementation, so Under the 256-bit memory access data structure, we tested the memory access latency of $stride=1,2,3,4,8,9,10,18$, and we can see that the buffer of the HLS buffer so that the Page Hit and Page Close latency of the memory access is not significantly different, but the primary latency data is not much distinctive, and the peak of the maximum latency has a significant shift with stride.

\begin{table}[!htbp]
	\centering
	\setlength{\abovecaptionskip}{+5pt}
	\setlength{\belowcaptionskip}{-12pt}
	\caption{latency and channel}
	\label{tahab:graphs}
	\begin{tabular}{|c|c|c|c|c|c|}
		\hline
		\multirowcell{2}{channel} &\multicolumn{2}{c|}{Minimums Latency} &\multicolumn{2}{c|}{Maximum Latency} &\multirowcell{2}{Average}  \\
		\cline{2-5}
		&MIN&AVG&MAX&AVG& \\
		\hline
		\hline
		0&58&   58.2391&   174&  106.7290&   60.5851\\
        \hline
       2&58&   58.2846&   174&  106.1408&   60.5856\\
       \hline
       4&58&   58.3060&   177&  106.1307&   60.5861\\
       \hline
       6&58&   58.2861&   174&  106.3653&   60.5860\\
       \hline
       8&58&   58.4740&   174&  104.2915&   60.5869\\
       \hline
       10&58&   58.2483&   176&  106.4918&   60.5850\\
       \hline
       12&71&   71.7729&   185&  157.5382&   74.1615\\
       \hline
       14&58&   58.3006&   175&  106.1307&   60.5859\\
       \hline
       16&58&   58.2501&   174&  106.3061&   60.5851\\
       \hline
       18&58&   58.2529&   175&  106.3207&   60.5850\\
       \hline
       20&58&   58.2723&   174&  106.2391&   60.5855\\
       \hline
       22&58&   58.2791&   174&  106.0310&   60.5852\\
       \hline
       24&58&   58.3348&   177&  105.4133&   60.5861\\
       \hline
       26&58&   58.2773&   174&  106.1195&   60.5854\\
       \hline
       28&58&   58.2276&   174&  106.6565&   60.5848\\
       \hline
       30&58&   58.2460&   176&  106.4774&   60.5850\\
       \hline
		\hline
	\end{tabular}
\end{table}



\subsection{Effect of Parameter} 
After the general latency benchmark, we will test some specific parameters based on the HLS implementation. The HLS high-level language features determine the influence of these parameters on memory access. Compared with the latency benchmark, the benchmark of these parameters will be more high-level HBM. Performance and characteristics under HLS implementation.

Combining high-level language features and HLS's access characteristics to the Axi bus system, we will test the memory access performance of HBM from the following four aspects:

1. Data structure benchmark based on memory access.

2. Memory burst based on the AXI bus structure 

3. Memory access test based on the outstanding data channel 

5. Memory access test based on HBM's address mapping policy (AMP).

We need to pay great attention to the memory subsytem on the FPGA, since memory transactions consume the majority of overall computing power and since the memory performance can be the bottleneck of overall performance. In the following, we quantitatively exiname the effect of each parameter related to the external memory instructions. 

In Xilinx FPGA implementation, the main memory access method for off-chip memory is AXI bus memory access, including AXI3, AXI4, and AXI stream. Generally speaking, AXI3 master memory access is the most common implementation method of the bus. Therefore, whether in RTL or HLS or even OpenCL, memory access performance is related to AXI related parameters. In HLS, the AXI memory access parameters include the explicit parameter Unit Size, and the implicit parameters Latency, Depth, Burst Size, and Outstanding Transactions. These four main parameters. Depth is a parameter that assumes the memory access depth and memory size, which is for simulation, so it does not affect actual performance.

\subsubsection{Effect of Latency} 


Latency is a parameter assumed for memory access delay, and its length will affect the length of the repeated implementation pipeline in the For Loop.

\subsubsection{Effect of Unit Size} 

Unit Size is an explicit parameter, and its impact on throughput is to show the characteristics of the data structure in the implementation. Generally speaking, since the primary storage unit of memory is bytes, to simplify and do representative experiments, we use 32bit \~ 512bit is the data structure of the benchmark (512bit is the maximum Unit Size of AXI single transmission). For comparison, we also test the int16 (structure), which shows in the X-axis of Figure~\ref{fig:unit_size}. The difference from 512bit is that one unit is 512bit, and the other is 16 int units.

Besides, show the resource consumption when varying Unit Size. 
As shown in Figure~\ref{fig:unit_size}, in the benchmark of data structure, as the length of unit size increases, the throughput of the memory access also improves, which is obviously with linear growth. However, the efficiency of using Arbitrary Integer Precision Types directly is not as good as using Data Structure; this is because of Arbitrary Integer Precision Types in HLS If disusing the bit range selection and bit operation, the system will construct an operation unit with the same length as the Integer. Compared with Structure, the length of the data unit inside the Data Structure define its operation unit, that is, if Arbitrary Integer Precision Types is 512bit and Structure is 16 Integer, then the system will set the operation length to 512bit for Arbitrary Integer, and Structure will have 16 independent Integer units, which results in higher performance in throughput.

As shown in Figure~\ref{fig:unit_size}, as lengthier Unit Size, the throughput is significantly improved, which conforms to the linear correlation.

\subsubsection{Effect of Burst Size} 

\begin{figure}[t]
\captionsetup[subfigure]{justification=centering}
\centering

\subfloat[ \phantom{HLSHLSHLHSLH       HLS.        HLSHHHHH} ]{%
  \includegraphics[width=2.5in]{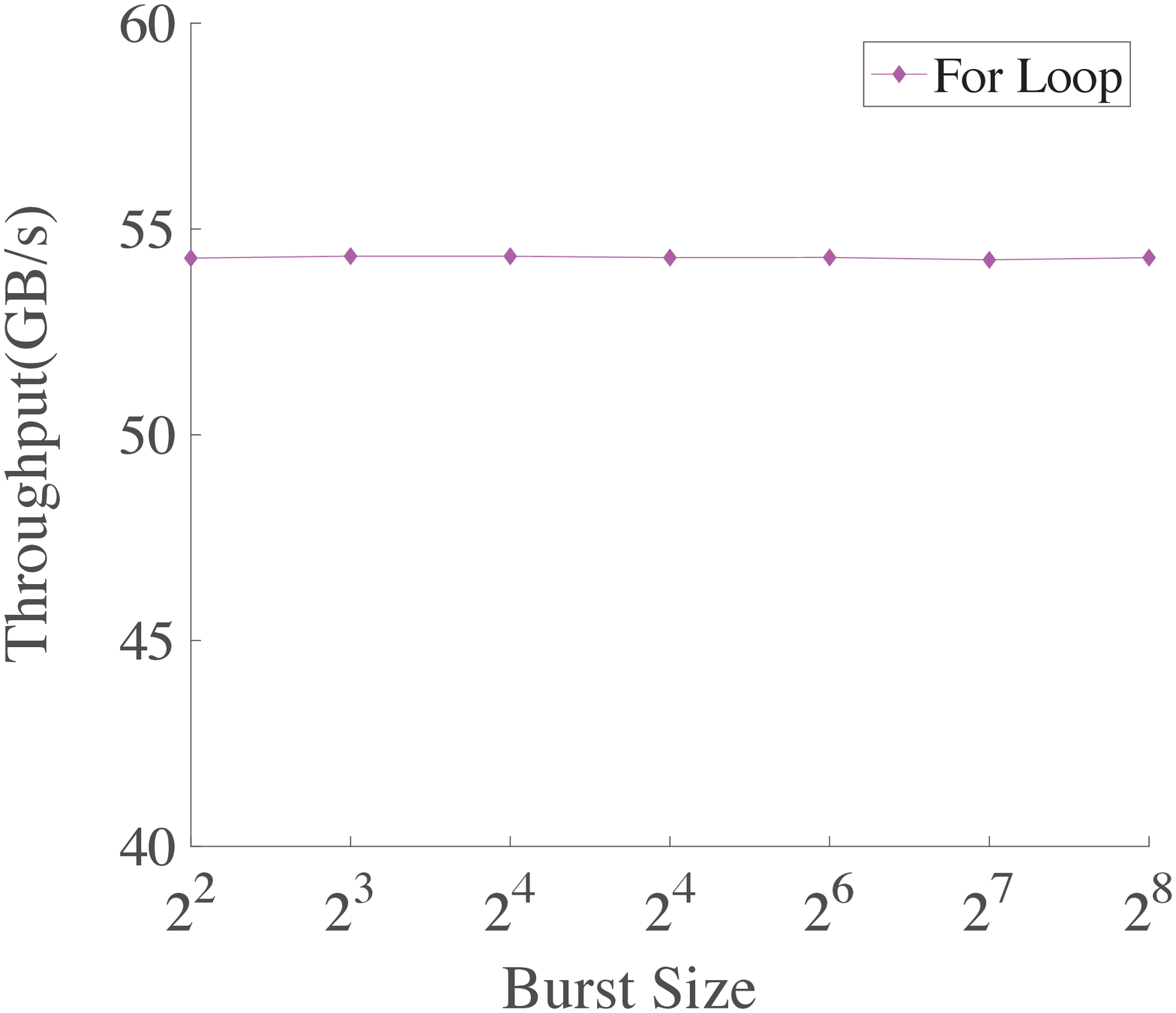}%
  \label{fig:burst_Pipeline}%
}

\subfloat[ \phantom{HLSHLSHLHSLH       HLS.        HLSHHHHH} ]{%
  \includegraphics[width=2.5in]{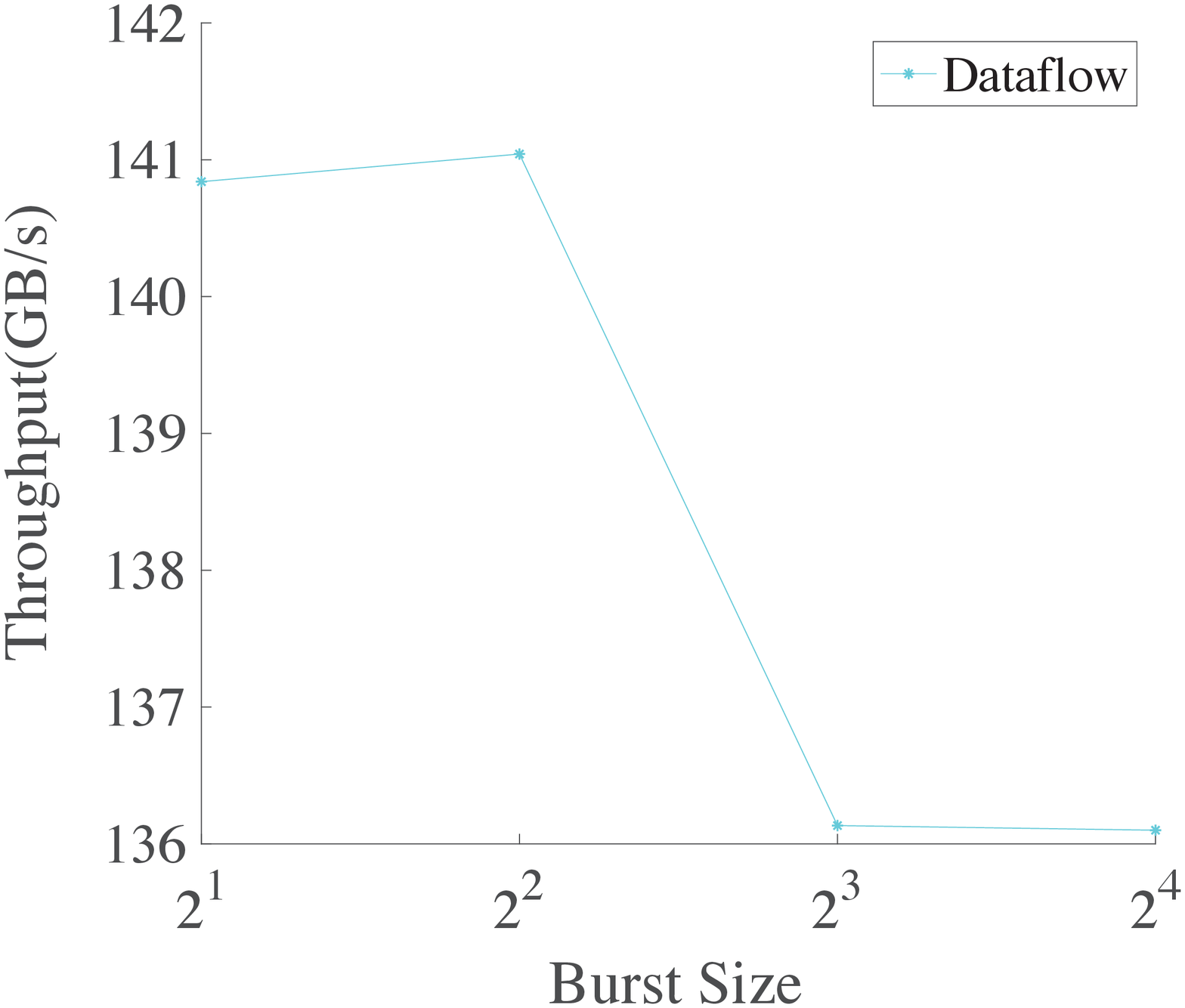}%
  \label{fig:burst_LOOP}%
}

\caption{Throughput With Burst Size}
\label{fig:burst_size}
\end{figure}

As shown in Figure~\ref{fig:burst_size}, from the architectural point of view, U280's access to HBM is wholly primary on the AXI bus structure, so the optimization of AXI memory access parameters can also optimize the HBM memory access performance. The implementation based on the Vitis platform (Whether implemented in HLS or RTL) There are restrictions on the access parameters of AXI (AXI4 master), and the main controllable parameter is the burst-related transmission signal. In the HLS implementation, we can easily set the burst length to test the seven transmission lengths that the burst is equal to 2, 4, 8, 16, 32, 64, 128, and 256.
\begin{table}[!htbp]
    \centering
    \setlength{\abovecaptionskip}{+5pt}
	\setlength{\belowcaptionskip}{-12pt}
    \caption{Utilization of Burst Size(Dataflow) Benchmark}
    \label{tab:UtilizationDataflowB}
    \begin{tabular}{|c|c|c|c|}
    \hline
    \multirowcell{2}{Burst\\ Size} &  \multicolumn{3}{c|}{Utilization}\\
     \cline{2-4}
    &  LUT& FF&BRAM\\
    \hline
        2  & 16.1\%  & 11.6\% & 11.4\%\\
        \hline
        4  & 16.3\%  & 11.7\% & 11.4\%\\
        \hline
        8  & 16.2\%  & 11.7\% & 11.4\%\\
        \hline
        16 & 16.2\%  & 11.7\% & 11.4\%\\
        \hline
    \end{tabular}
\end{table}

As shown in Figure~\ref{fig:burst_size}, the burst size has a limited impact on the continuous For Loop and Dataflow memory access implementation, but from the Table~\ref{tab:UtilizationDataflowB} and Table~\ref{tab:UtilizationLOOPB}, the BRAM consumption increases as the burst size increases. Therefore, the burst size has less impact on the HLS implementation that can issue a transmission request every cycle.

\begin{table}[!htbp]
    \centering
    \setlength{\abovecaptionskip}{+5pt}
	\setlength{\belowcaptionskip}{-12pt}
    \caption{Utilization of Burst Size(Loop) Benchmark}
    \label{tab:UtilizationLOOPB}
    \begin{tabular}{|c|c|c|c|}
    \hline
    \multirowcell{2}{Burst\\Size} &  \multicolumn{3}{c|}{Utilization}\\
     \cline{2-4}
    &   LUT& FF&BRAM\\
    \hline
    2  & 21.8\%  & 23.7\%  & 37.4\%\\
    \hline
    4  & 21.8\%  & 23.7\%  & 37.4\%\\
    \hline
    8  & 21.8\%  & 23.7\%  & 37.4\%\\
    \hline
    16 & 21.8\%  & 23.7\%  & 37.4\%\\
    \hline
    32 & 21.7\%  & 23.7\%  & 37.4\%\\
    \hline
    64 & 21.6\%  & 23.6\%  & 49.8\%\\
    \hline

    \end{tabular}
\end{table}

\subsubsection{Effect of Outstanding Transactions} 

AXI-based memory access behavior, in addition to the explicit burst parameter, there is an implicit outstanding memory access parameter, which sets the number of cache channels to cache multiple requests of memory access in parallel, which can alleviate memory access performance degradation caused by memory access latency. The Outstanding parameter is transparent in the implementation of RTL; as shown in Figure~\ref{fig:outstanding}, its implementation is transparent to the AXI port parameters. However, for the HLS implementation, this parameter can be set and tested. According to its characteristics, we have the corresponding parameters for outstanding benchmark, including 4, 8, 16, 32, 64, 128, 256, and other six transmission channel parameters for the benchmark.
\begin{table}[!htbp]
    \centering
    \setlength{\abovecaptionskip}{+5pt}
	\setlength{\belowcaptionskip}{-12pt}
    \caption{Utilization of Outstanding Transaction(Dataflow) Benchmark}
    \label{tab:UtilizationOutstandingD}
    \begin{tabular}{|c|c|c|c|}
    \hline
    \multirowcell{2}{Outstanding\\Transaction}& \multicolumn{3}{c|}{Utilization}\\
     \cline{2-4}
    &   LUT& FF&BRAM\\
    \hline
        2  & 21.8\%  & 22.1\%  & 37.4\%\\
        \hline
        4  & 21.9\%  & 22.1\%  & 37.4\%\\
        \hline
        8  & 21.9\%  & 22.1\%  & 37.4\%\\
        \hline
        16 & 21.9\%  & 22.1\%  & 37.4\%\\
        \hline
        32 & 21.9\%  & 22.1\%  & 37.4\%\\
        \hline
        64 & 22.0\%  & 22.1\%  & 45.6\%\\
        \hline
    \end{tabular}
\end{table}
As shown in Figure~\ref{fig:outstanding}, the impact of Outstanding Transactions on Dataflow and For Loop is quite apparent, especially Dataflow, which is in line with linear growth. However,  as shown in Table~\ref{tab:UtilizationOutstandingD}, the growth of Outstanding Transactions is also positively correlated with BRAM consumption, But because BRAM bandwidth and depth are discrete, the change is not significant.

\subsubsection{Effect of Stride} 

In the stride benchmark, we use the standard for Loop to test, such as Algorithm~\ref{alg:Stride memory access}, we used the Host to transport the address offset parameter to the implementation on board. then
we use the standard For Loop to test the performance on different strides, which we consider how to influence throughput,  so we do not limit burst and outstanding in this benchmark.

In the sequential read and write benchmark, we also test it with a standard For Loop. Also, to obtain the highest performance, we do not limit burst and outstanding and restrict the stride to 1. Simultaneously, to obtain its performance in reading, writing, and reading and writing, we will also test these three states in sequence.

\subsubsection{Effect of number of kernels}

The number of kernels in HLS is also an interesting parameter that has an impact on bandwidth. Under the same memory channel usage (32 channels), the impact of different kernel numbers on Throughput decreases as the kernel increases. As shown in the Table~\ref{tab:KernelThroughput}, we tested 1, 2, 4, 8, 16, 32 kernels, of which 1, 2 kernels use Dataflow, and the others use For Loop. It also shows that the performance is better when there are fewer kernels.
\begin{table}[!htbp]
    \centering
    \setlength{\abovecaptionskip}{+5pt}
	\setlength{\belowcaptionskip}{-12pt}
    \caption{Throughput Benchmark of Number of Kernels}
    \label{tab:KernelThroughput}
    \begin{tabular}{|c|c|c|c|c|}
    \hline
    \multirowcell{2}{Number \\of Kernel}& \multirowcell{2}{Throughput\\ (GB/s)} &  \multicolumn{3}{c|}{Utilization}\\
     \cline{3-5}
    & &  LUT& FF&BRAM\\
    \hline
      32 kernel & 136.093 & 21.6\%  &11.9\%&36.6\%\\
     \hline
      16 kernel& 187.093 &21.8\% &23.7\%&37.4\%\\
     \hline
      8 kernel& 373.381 &20.5\%&22.8\%&37.4\%\\
     \hline
      4 kernel& 406.523 &21.9\%&21.8\%&37.4\%\\
     \hline
      2 kernel$^*$& 418.116 &16.5\%&11.9\%&11.8\%\\
     \hline
      1 kernel$^*$& 421.691 &16.1\%&11.6\%&11.4\%\\
     \hline
    \end{tabular}
\end{table}

\subsubsection{OLD!!!!}

\begin{figure}[!htbp]
	\centering
	\setlength{\abovecaptionskip}{+5pt}
	\setlength{\belowcaptionskip}{-15pt}
    \includegraphics[width = 2.5in]{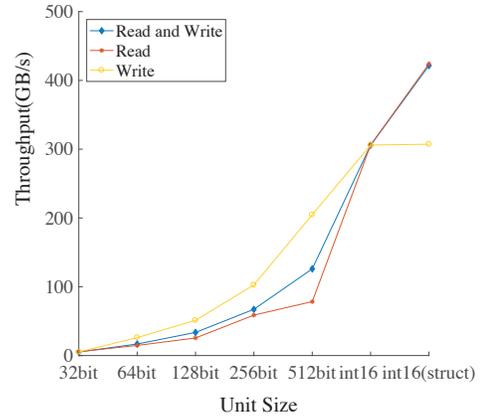}

	\caption{Throughput With Various Unit Size}
	\label{fig:unit_size}
\end{figure}

\begin{figure}[htbp]
\includegraphics[width=2in]{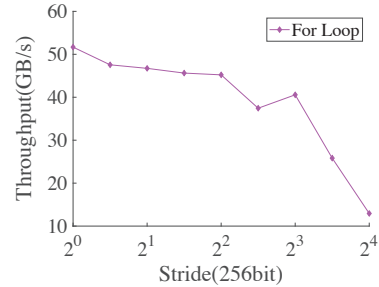}
\caption{Throughput With Various Stride (Loop)}
\label{fig:stride_pipe}%
\end{figure}

\begin{figure}[htbp]
\includegraphics[width=2in]{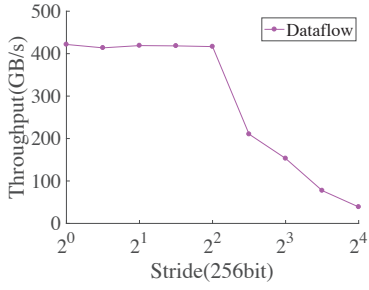}
\caption{Throughput With Various Stride (Dataflow)}
\label{fig:stride_loop}%
\end{figure}




In the random performance benchmark, to cover the impact of different random addresses on the memory access performance, we use two different address generation methods for memory access to test the random performance: random generator and point chasing\cite{weisz2016study}. Random generator refers to directly using to generate random addresses for memory access on board. Since there is no standard random number generation library for HLS-based implementations, we use line feedback shift register (LSFR)~\cite{krawczyk1994lfsr,panda2012fpga,schellekens2006fpga,tsoi2003compact} for random number generation, such as Algorithm~\ref{alg:LFSR} and Algorithm~\ref{alg:point_chasing}, due to the random address generated independently, we use a pipeline to conduct this random performance test. Point chasing uses random numbers to generate a random linked list, stores the linked list in HBM (Host completes this part), and then uses the random linked list in the HLS implementation to obtain random performance.

\begin{algorithm}[htbp]
    \caption{LFSR random generate}
    \Input{The seed data $SEED$,The LSFR initial signal $START$, Maximal-length polynomials for LSFR $x^{e_0}+x^{e_1}+...+x^{e_{k}}+1$}
    \Output{the current random number $LFSR$}
    \label{alg:LFSR}
    \If{$START$ is equal to $1$}{
        $LSFR \gets SEED$;
    }
    \Else{
        \ForEach{$j \in [0,k]$}
        {
            $BITS \gets BITS \And LSFR.Getbit(e_0-e_j)$;\\
        }
        $LSFR >> 1$;\\
        $LSFR.Setbit(e_0,BITS);$
    }
\end{algorithm}

As shown in Figure~\ref{fig:unit_size}, in the benchmark of data structure, as the length of unit size increases, the throughput of the memory access also improves, which is obviously with linear growth. However, the efficiency of using Arbitrary Integer Precision Types directly is not as good as using Data Struct; this is because of Arbitrary Integer Precision Types in HLS If disusing the bit range selection and bit operation, the system will construct an operation unit with the same length as the Integer. Compared with Struct, the length of the data unit inside the Data Struct define its operation unit, that is, if Arbitrary Integer Precision Types is 512bit and Struct is 16 Integer, then the system will set the operation length to 512bit for Arbitrary Integer, and Struct will have 16 independent Integer units, which results in higher performance in throughput.

\begin{algorithm}[htbp]
    \caption{Point chasing address benchmark}
    \Input{The total transaction $I$, memory data channel ${H_n}$}
    \Output{memory data out $D$,memory access address$ADDR$}
    \label{alg:point_chasing}
    $ADDR \gets 0$
    \ForEach{$i \in [0,I-1]$}{
        $D\gets {H_n}[ADDR]$;\\
        $ADDR \gets D$;
    }
\end{algorithm}





\section{Optimizing Memory Access Pattern}

The main goal of throughput testing is to examine the exact relationship between throughput and parameter 
\& resource consumption. For each load/store site, such a relationship can guide the FPGA programmer who intends to use Vitis (generally HLS) to choose the right optimization level that not only meets throughput requirement but also consumes as few resources as possible.   

In the benchmark phase, to reduce the impact of actual implementation on performance, we will test burst and outstanding benchmark under the two architectures of standard For Loop and Dataflow at the same time. For the standard data structure benchmark, since it has nothing to do with implementation, it will only be tested in a standard For Loop. For dependency false, we use Dataflow for this kind of test with a closer relationship with the HLS implementation, which has a relatively high degree of parallelism. 

After completing the standard benchmark, we consider the impact of some actual implementations on performance. Because of the above benchmarks on latency and some parameters, to shield the performance impact of address jumps, all we consider are address steps Sequential read/write is 1. So next, we will perform multiple tests on the memory access performance of multiple different step sizes, standard sequential read/write, random read/write, and other conventional applications such as AI and other HPC architectures in HBM.

\begin{algorithm}[h]
    \caption{Stride Address Benchmark}
    \Input{The total transaction $I$,Work group size of HBM channel $G$,The stride of address $S$, HBM data channel ${H_n}$}
    \Output{HBM data out $D$,HBM access address$ADDR$}
    \label{alg:Stride memory access}
    $ADDR \gets 0$
    \ForEach{$i \in [0,I-1]$}{
        $D\gets {H_n}[\left(ADDR+S\right) \mod G]$;
    }
\end{algorithm}

As shown in Figure~\ref{fig:stride_pipe} and Figure~\ref{fig:stride_loop} of the benchmark on stride, we find that as stride increases, the memory access performance is a significant reduction, this is due to the outstanding failure to bridge the memory access latency, which is the same as the conclusion we found in the random access test, As shown in the Table~\ref{tab:Throughput}.

\begin{table}[!htbp]
    \centering
    \setlength{\abovecaptionskip}{+5pt}
	\setlength{\belowcaptionskip}{-12pt}
    \caption{Throughput Benchmark of Random (LFSR)}
    \label{tab:ThroughputLFSR}
    \begin{tabular}{|c|c|c|c|c|}
    \hline
    \multirowcell{2}{Outsanding\\ Transactions}& \multirowcell{2}{Throughput\\ (GB/s)} &  \multicolumn{3}{c|}{Utilization}\\
     \cline{3-5}
    & &  LUT& FF&BRAM\\
    \hline
     2 & 5.96 & 16.1\%& 11.6\%& 11.4\%\\
     \hline
     4 & 6.00 & 16.1\%& 11.7\%& 11.4\%\\
     \hline
     8 & 5.57 & 16.2\%& 11.8\%& 11.4\%\\
     \hline
    16 & 5.82 & 21.8\%& 23.7\%& 37.4\%\\
    \hline
    32 & 5.57 & 20.9\%& 23.5\%& 36.6\%\\
    \hline
    64 & 5.64 & 20.8\%& 23.5\%& 36.6\%\\
     \hline
    \end{tabular}
\end{table}
\begin{table}[!htbp]
    \centering
    \setlength{\abovecaptionskip}{+5pt}
	\setlength{\belowcaptionskip}{-12pt}
    \caption{Throughput Benchmark of Random Comparison}
    \label{tab:ThroughputRandom}
    \begin{tabular}{|c|c|c|c|c|}
    \hline
    \multirowcell{2}{Benchmark\\Model}& \multirowcell{2}{Throughput\\ (GB/s)} &  \multicolumn{3}{c|}{Utilization}\\
     \cline{3-5}
    & &  FF&LUT &BRAM\\
    \hline
      Sequential & 421.68 & 16.1\%&11.6\%&11.4\%\\
     \hline
      random(LFSR)& 5.82 &16.2\%& 11.8\%& 11.4\%\\
     \hline
      random(Point)& 0.994161 &16.5\%&11.5\%&11.4\%\\
     \hline
    \end{tabular}
    
\end{table}

\section{Optimizing Memory Access Pattern for Applications}
The performance of memory-bound application is highly sensitive to its underlying memory access pattern. 

\subsection{Machine Learning Inference}

We present convolutional network computing efficiency as a fundamental benchmark for machine learning implementation, which performs a convolution calculation with a kernel of $11*11$ on a $1920*1080$ matrix in this benchmark. In the implementation, we utilize three types of implementation 1. CPU-based benchmark, 2. HBM-based single-kernel with the dual-channel benchmark (read and write on different memory channels) 3. HBM-based 32-channel  with 16 kernels benchmark (each kernel contains one read and another write channels).
As shown in the Table~\ref{tab:Convolution}, the runtime is that FPGA implementation can achieve a performance improvement of more than $100X$ the implementation for CPU, and the parallel performance of multi-channel is also $10X$ higher than the performance of a single kernel, and its resource consumption is only about $2X$.

\subsection{Database (DB)}
It is well known that evaluating a database query, consisting of a few database operators, is typically memory-bound on modern hardware like CPUs, so it is critical to understand and optimize the memory performance of each database operator so as to achieve high overall performance. Accordingly, the database community presents a few basic memory access patterns, upon which memory access costs of these database operators are modeled~\cite{manegold2002generic}. The performance characteristics of these basic patterns, together with the corresponding optimizations, are well analyzed on modern CPUs. However, these basic patterns are still not systematically analyzed on FPGAs. So we analyze four basic patterns include repetitive sequential traversal (\emph{rs\_stra}), repetitive random traversal (\emph{rr\_stra}) and random access (\emph{r\_acc}), and interleaved multi-cursor sequential access (\emph{nest}).\footnote{The other two basic patterns are subset of the above four patterns. }
Table~\ref{tab:database} illustrates the throughput of each basic pattern, as well as the corresponding FPGA resource consumption. 

\noindent {\bf Optimizing \emph{rs\_stra}. } Larger unit size leads to higher memory throughput. Larger stride leads to lower memory throughput, while large unit size can amortize memory throughput loss. 

\noindent {\bf Optimizing \emph{rr\_stra}. } Larger unit size leads to higher memory throughput. 

\noindent {\bf Optimizing \emph{r\_acc}. } Larger unit size leads to higher memory throughput. 

\noindent {\bf Optimizing \emph{nest}. } Larger unit size and/or appropriate stride leads to high memory throughput.


\begin{table}[!htbp]
    \centering
    \setlength{\abovecaptionskip}{+5pt}
	\setlength{\belowcaptionskip}{-12pt}
    \caption{Database Benchmark}
    \label{tab:database}
    \begin{tabular}{|c|c|c|c|c|}
    \hline
    
    \multirowcell{2}{Access\\Patterns}& \multirowcell{2}{Throughput\\ (GB/s)} &  \multicolumn{3}{c|}{Utilization}\\
     \cline{3-5}
    & &  LUT& FF&BRAM\\
    
    \hline
    rs\_tra & 13.26 &9.0\% & 7.1\% & 10.8\%\\
     \hline
     rr\_tra & 3.51&9.1\%& 7.1\% & 11.2\%\\
     \hline
     r\_acc &0.68 & 9.1\%& 7.1\%&10.7\%\\
     \hline
     nest & 421.89 &16.5\%  & 11.9\% & 11.8\%\\
 \hline
\end{tabular}
\end{table}

\begin{table}[!htbp]
    \centering
    \setlength{\abovecaptionskip}{+5pt}
	\setlength{\belowcaptionskip}{-12pt}
    \caption{Convolution Benchmark}
    \label{tab:Convolution}
    \begin{tabular}{|c|c|c|c|c|c|c|}
    \hline
    \multirowcell{2}{Type}&
    \multirowcell{2}{Channel}&
    \multirowcell{2}{$BW$\\(GB/s)} & 
    \multirowcell{2}{$T$\\(s)} & 
    \multicolumn{3}{c|}{Utilization}\\
     \cline{5-7}
    & &  & &FF&LUT &BRAM\\
    
    \hline

    CPU & - & 0.263 & 0.06 & - & - & - \\
    \hline
    FPGA & 2 & 0.0080 & 2.04 & 9.1\% & 7.0\% & 11.3\%\\
    \hline
    FPGA & 32 & 0.012 & 21.0 & 18.8\% & 14.8\% & 31.3\%\\
     \hline
\end{tabular}
\end{table}

\section{Conclusion}

By benchmark the memory access bandwidth of the Vitis platform's FPGA platform implementation on the HLS, the HLS on the Vitis platform can fully present to the performance of HBM and other storage systems, especially for HBM. Compared to the direct use of Vivado , our architecture can make more comfortable reach the peak of theoretical performance. Furthermore, for HLS, we can also make a more detailed and actual performance benchmark on HBM from the realistic system's perspective. In this benchmark, we conducted a comprehensive examination on the HBM2 of all two stacks and its 32 pseudo channels, and obtained their memory access characteristics, and compared them with the benchmark of Shuihai. Accordingly, we propose an architecture for FPGA memory access performance and memory access system under the Vitis platform that is easy to test and expand. This architecture provides a comprehensive and detailed overview of the off-chip memory structure of FPGA, mainly HBM. The benchmark includes latency, bandwidth throughput, random and continuous performance, particular parameters, and some tests at the system (such as database) level, so that we can understand the feature of FPGA memory access, especially HBM2, under the Vitis platform and HLS implementation, And its advantages with CPU/GPU implementation. We will extend the benchmark implemented in this paper to different FPGA development boards and provide open source code for more benchmark.

\section*{Acknowledgment}



\bibliographystyle{ACM-Reference-Format}
\balance
\bibliography{myref}

\end{document}